\documentclass[twocolumn,showpacs,prr,floatfix,amsmath,amssymb,notitlepage,superscriptaddress]{revtex4-1}
\usepackage{epsfig}
\usepackage{subfigure}
\usepackage{amsmath}
\usepackage{color}
\usepackage{amssymb}
\usepackage{setspace}
\usepackage{graphicx}
\usepackage{dcolumn}
\usepackage{bm}
\usepackage{times}
\usepackage{enumerate}
\usepackage{footnote}
\input epsf

\graphicspath{{figures/}}

\begin{document}

\author{Per Sebastian Skardal}
\email{persebastian.skardal@trincoll.edu} 
\affiliation{Department of Mathematics, Trinity College, Hartford, CT 06106, USA}

\author{Llu\'{i}s Arola-Fern\'{a}ndez}
\affiliation{Departament d'Enginyeria Inform\`{a}tica i Matem\'{a}tiques, Universitat Rovira i Virgili, 43007 Tarragona, Spain}

\author{Dane Taylor}
\affiliation{Department of Mathematics, University at Buffalo, State University of New York, Buffalo, NY 14260, USA}

\author{Alex Arenas}
\affiliation{Departament d'Enginyeria Inform\`{a}tica i Matem\'{a}tiques, Universitat Rovira i Virgili, 43007 Tarragona, Spain}

\title{Higher-order interactions can better optimize network synchronization}

\begin{abstract}
Collective behavior plays a key role in the function of a wide range of physical, biological, and neurological systems where empirical evidence has recently uncovered the prevalence of higher-order interactions, i.e., structures that represent interactions between more than just two individual units, in complex network structures. Here, we study the optimization of collective behavior in networks with higher-order interactions encoded in clique complexes. Our approach involves adapting the Synchrony Alignment Function framework to a new \emph{composite Laplacian matrix} that encodes multi-order interactions including, e.g., both dyadic and triadic couplings. We show that as higher-order coupling interactions are equitably strengthened, so that overall coupling is conserved, the optimal collective behavior improves. We find that this phenomenon stems from the broadening of a composite Laplacian's eigenvalue spectrum, which improves the optimal collective behavior and widens the range of possible behaviors. Moreover, we find in constrained optimization scenarios that a nontrivial, ideal balance between the relative strengths of pair-wise and higher-order interactions leads to the strongest collective behavior supported by a network. This work provides insight into how systems balance interactions of different types to optimize or broaden their dynamical range of behavior, especially for self-regulating systems like the brain.
\end{abstract}


\maketitle

\section{Introduction}
Complex networks provide the structural architecture for dynamical processes from a wide array of disciplines, and therefore their study constitutes an important fundamental area of research in physics, mathematics, biology, and engineering~\cite{Strogatz2003,Pikovsky2003,Arenas2008PhysRep}. Collective behaviors, i.e., consensus and synchronization, play particularly critical roles in the functionality of systems in many applications, with recent interest paid to applications including brain oscillations~\cite{Schnitzler2005Nature,Deco2011Frontiers,Fell2011Nature}, cell signaling~\cite{Prindle2012Nature,Prindle2014Nature}, and power grids~\cite{Rohden2012PRL,Skardal2015SciAdv}. Moreover, various combinations of local dynamics with different microscopic and macroscopic topological network properties have been shown to give rise to a wide range of novel collective behaviors, including explosive synchronization transitions~\cite{GomezGardenes2011PRL,Skardal2014PRE}, chimera states~\cite{Panaggio2015Nonlinearity,Nicolaou2019PRX}, and macroscopic chaos~\cite{Skardal2015PRE,Bick2018Chaos}, thus having important effects on system functions. 

In addition to typical pairwise/dyadic interactions in network-coupled systems, recent work points to the presence of higher-order, e.g., triadic, interactions in both brain networks~\cite{Yu2011Neuro,Petri2014Interface,Giusti2016JCN,Reimann2017,Sizemore2018JCN} and generic limit-cycle oscillator systems~\cite{Ashwin2016PhysD,Leon2019PRE}. The presence of such interactions is often encoded in simplicial complexes or hypergraphs~\cite{Horak2009,Salnikov2019EJP,Schaub2020SIAM,Battiston2020PhysRep} and has prompted the network science community to develop tools to better understand the impact of such higher-order interactions on collective dynamics. To date, a handful of studies have explored the role of higher-order interactions in collective dynamics in heterogeneous systems~\cite{Tanaka2011PRL,Bick2016Chaos,Skardal2019PRL,Millan2020PRL,Skardal2020A,Skardal2020B,Lucas2020PRR,Iacopini2019NatComms}, but unlike real naturally-occurring or engineered systems that are often optimized for a particular task, these initial studies tend to utilize random configurations or mean-field assumptions. At present, the role of higher-order interactions for optimized systems is largely unknown for collective dynamics and other dynamical processes.

In this paper, we study collective dynamics in networks with higher-order interactions, focusing on collective behavior in optimized systems. 
To quantify the optimal collective behavior supported by a given network structure with higher-order interactions, we introduce a \emph{composite Laplacian matrix}, which encodes the collective dynamics and network structure at multiple orders in a weighted simplicial complex and generalizes the Synchrony Alignment Function (SAF) framework~\cite{Skardal2014PRL} to this case. For the case of simple dyadic interactions, the SAF has been used to uncover the critical properties needed to optimize collective behavior in networks with heterogeneous dynamics and has proven to be flexibly adaptable to a wide range of realistic constraints and scenarios~\cite{Skardal2016Chaos,Taylor2016SIAM,Skardal2017Chaos,Skardal2019SIAM,Arola2021Chaos}, as it encodes the interplay between heterogeneous dynamical units and heterogeneous network structure. We emphasize that in this context optimal refers to a system being as strongly synchronized as possible. 

Generalizing the SAF framework, applying it in this new context, and analyzing the spectral properties of the composite Laplacian reveals important new properties of networks with higher-order interactions. Specifically, as higher-order interactions are strengthened in a system at the expense of weakening dyadic interactions to conserve the total coupling in the network, the eigenvalue spectrum of a composite Laplacian broadens. This, in turn, increases the dominant eigenvalue, which is key to improving the optimal state supported by the network. Complementing this improvement of the optimal collective state, the broadening of the eigenvalue spectrum also increases the overall range of possible states. This phenomenon contrasts sharply with synchronization of identical oscillators, where optimization of a network for identical synchronization reduces to a contraction of the eigenvalue spectrum~\cite{Barahona2002PRL,Nishikawa2017PRX}. We close by exploring a realistic constrained optimization problem where local dynamics are not freely tunable, but must be allocated from a pre-defined set, and it is revealed that a network's ideal configuration is realized by a nontrivial, critical balance between the strength of dyadic and triadic interactions.

\section{Dynamics and Modeling}

We begin with a higher-order generalization of the Kuramoto model~\cite{Skardal2020A,Kuramoto1984} that consists of $N$ phase oscillators whose states $\theta_i$, for $i=1,\dots,N$, evolve according to 
\begin{align}
\dot{\theta}_i = \omega_i &+ \frac{K_1}{\langle k^{(1)}\rangle}\sum_{j=1}^NA_{ij}\sin\left(\theta_{j}-\theta_i\right)\nonumber\\
&+\frac{K_2}{2\langle k^{(2)}\rangle}\sum_{j=1}^N\sum_{l=1}^NB_{ijl}\sin(2\theta_j-\theta_l-\theta_i).\label{eq:01}
\end{align}
Here, $\omega_i$ is the natural frequency of oscillator $i$, $K_1$ and $K_2$ are coupling strengths that are associated with 1- and 2-simplex interactions, respectively, $A$ is a 1-simplex adjacency matrix, and $B$ is a 2-simplex adjacency tensor. We assume the network to be unweighted and undirected so that $A_{ij}=A_{ji}=1$ if and only if a link exists between oscillators $i$ and $j$, and $B_{ijl}=B_{ilj}=B_{jil}=B_{jli}=B_{lij}=B_{lji}=1$ if and only if a triadic interaction exists between oscillators $i$, $j$ and $l$. 

While the 1- and 2-simplex coupling topologies may in general be uncorrelated for the case of a general hypergraph, here we assume the system corresponds to a simplicial complex so that the existence of a triadic interaction $(i,j,l)$ requires the existence of dyadic interactions $(i,j)$, $(j,l)$ and $(l,i)$. (Formally, the boundary of any 2-simplex in the simplicial complex must also be contained in the simplicial complex.) Moreover, we restrict our attention here to `clique complexes'~\cite{Kahle2009} in which all triangles give rise to 2-simplices, which allows the 3-tensor $B$ to be completely determined by matrix $A$, i.e., $B_{ijl} = A_{ij}A_{jl}A_{li}$. 

The respective coupling strengths in Eq.~(\ref{eq:01}) are scaled by the 1- and 2-simplex mean degrees $\langle k^{(1)}\rangle$ and $\langle k^{(2)}\rangle$, which are population averages of the 1- and 2-simplex degrees $k_i^{(1)}=\sum_{j=1}^NA_{ij}$ and $k_i^{(2)}=\frac{1}{2}\sum_{j=1}^N\sum_{l=1}^NB_{ijl}$. This scaling ensures that the overall connectivity is maintained between the 1- and 2- simplex structure. In other words, by conserving the sum $K = K_1+K_2$, we fix the overall amount of coupling in the network, regardless of the specific topologies encoded in $A$ and $B$. To this end, we introduce a new \emph{bias parameter} $\alpha\in[0,1]$, defined via $K_1=(1-\alpha)K$ and $K_2=\alpha K$, so that $\alpha\approx0$ corresponds to a system where $1$-simplex interactions are stronger than $2$-simplex interactions and vice-versa if $\alpha \approx 1$. 

\begin{figure}[t]
\centering
\epsfig{file =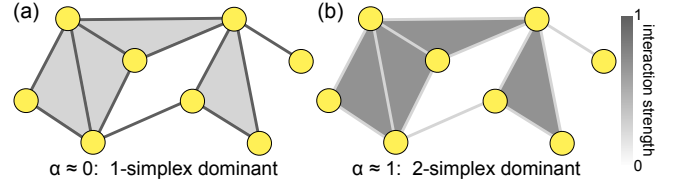, clip =,width=1.0\linewidth }
\caption{\emph{Weighted simplicial complexes encode the balancing of multi-order interactions.} An illustration of a small network with (a) 1- vs (b) 2-simplex dominated coupling. Shading indicates the relative strength of dyadic and triadic interactions which after rescaling by the respective mean degree, conserve the total coupling strength such that the \emph{bias parameter} $\alpha$ equitably tunes 1- vs 2-simplex interactions.} \label{fig1}
\end{figure}

Examples of a small toy network with 1- and 2-simplex dominated coupling are illustrated in Fig.~\ref{fig1} (a) and (b), respectively, where dyadic and triadic interactions are shaded to denote relative interaction strengths. Lastly, we note that other higher-order interaction terms may exist in other formulations of a higher-order Kuramoto model~\cite{Ashwin2016PhysD,Leon2019PRE}. We find that these yield qualitatively similar results as what is presented below, and so we focus our attention on the combination of 1- and 2- simplex interactions in Eq.~(\ref{eq:01}), and address additional triadic coupling terms in Appendix~\ref{appA}.

\section{Optimized Systems}

Since our focus here is on optimization, which again we emphasize refers to a system being as strongly synchronized as possible, we consider the strongly synchronized regime where $|\theta_j-\theta_i|\ll1$, allowing us to linearize Eq.~(\ref{eq:01}) to
\begin{align}
\dot{\theta}_i\approx\omega_i &- K\left[(1-\alpha)\left(k_i^{(i)}\theta_i-\sum_{j=1}^NA_{ij}\theta_j\right)/\langle k^{(1)}\rangle\right.\nonumber\\
&+\alpha\left(k_i^{(2)}\theta_i-\sum_{j=1}^NA_{ij}\left(\sum_{l=1}^NA_{jl}A_{li}\right)\theta_j\right.\nonumber\\
&~~~~~~\left.\left.+\frac{1}{2}\sum_{j=1}^NA_{ji}\left(\sum_{l=1}^NA_{il}A_{lj}\right)\theta_j\right)/\langle k^{(2)}\rangle\right],\label{eq:02}
\end{align}
or in vector form,
\begin{align}
\dot{\bm{\theta}}=\bm{\omega} - KL\bm{\theta},\label{eq:03}
\end{align}
where $L=(1-\alpha)L^{(1)} + \alpha L^{(2)}$ is a composite Laplacian that is a weighted average of the first and second-order Laplacians, which we define, respectively as $L^{(1)}=(D^{(1)}-A^{(1)})/\langle k^{(1)}\rangle$ and $L^{(2)}=(D^{(2)}-\left(A^{(2)}-A^{(2)T}/2\right))/\langle k^{(2)}\rangle$. The matrix $L^{(1)}$ is simply a scaled version of the typical combinatorial Laplacian with $D^{(1)}=\text{diag}(k_1^{(1)},\dots,k_N^{(1)})$ and $A^{(1)}=A$, while $L^{(2)}$ encodes the 2-simplex interactions with $D^{(2)}=\text{diag}(k_1^{(2)},\dots,k_N^{(2)})$ and $A^{(2)}=A*(A^2)^T$, where $*$ represents the Hadamard (i.e., element-wise) product. In addition to serving as the linear approximation of the nonlinear dynamics given in Eq.~(\ref{eq:01}), Eqs.~(\ref{eq:02}) and (\ref{eq:03}) also describe a forced consensus dynamics on the same network structure with higher-order interactions. Optimizing the synchronization dynamics using the linear approximation is equivalent to optimizing the consensus dynamics.

To optimize Eqs.~(\ref{eq:01})--(\ref{eq:03}) we enter the rotating reference frame $\theta\mapsto\theta+\langle \omega\rangle t$ (which allows us to effectively set the mean frequency to zero in both the nonlinear and linear dynamics), and we search for fixed points. Applying the Moore-Penrose pseudoinverse of the composite Laplacian~\cite{BenIsrael}, $L^\dagger=\sum_{j=2}^N\lambda_j^{-1}\bm{v}^j\bm{v}^{jT}$, where $0=\lambda_1<\lambda_2\le\cdots\le\lambda_N$ are the eigenvalues of $L$ and its eigenvectors $\{\bm{v}^j\}_{j=1}^N$ form an orthonormal basis for $\mathbb{R}^N$, yields the fixed point
\begin{align}
\bm{\theta}^*=\frac{L^\dagger\bm{\omega}}{K}.\label{eq:04}
\end{align}
From the viewpoint of consensus dynamics, the degree of consensus may be evaluated directly by the variance of the fixed point, $\|\bm{\theta}^*\|^2/N$. On the other hand, the degree of synchronization in the higher-order Kuramoto model is given by the magnitude $r$ of the order parameter $z=re^{i\psi}=N^{-1}
\sum_{j=1}^Ne^{i\theta_j}$, which represents the centroid of all oscillators when placed on the complex unit circle. To leading order, the degree of synchronization of the fixed point is $r\approx1-\|\bm{\theta}^*\|^2/2N$. Thus, consensus and synchronization dynamics are both optimized by minimizing the variance of the fixed point, $\|\bm{\theta}^*\|^2/N$. Using the form of $L^\dagger$ given above and that $\|\bm{\theta}^*\|^2=\langle\bm{\theta}^*,\bm{\theta}^*\rangle$, we have that
\begin{align}
\frac{\|\bm{\theta}^*\|^2}{N}=\frac{J(\bm{\omega},L)}{K^2},~~\text{where}~~J(\bm{\omega},L)=\frac{1}{N}\sum_{j=2}^N\frac{\langle\bm{v}^j,\bm{\omega}\rangle^2}{\lambda_j^2}.\label{eq:05}
\end{align}

The function $J(\bm{\omega},L)$ is known as the Synchrony Alignment Function (SAF), which was first introduced in Ref.~\cite{Skardal2014PRL} in the context of an objective function for optimizing the synchronization properties of a network of heterogeneous oscillators. Minimizing $J(\bm{\omega})$ serves to optimize $\|\bm{\theta}^*\|^2/N$ and $r$ and can be explored under a wide variety of constraints~\cite{Skardal2016Chaos,Taylor2016SIAM,Skardal2017Chaos,Skardal2019SIAM,Arola2021Chaos}. Inspecting the contributions to the SAF, we note that each term corresponds to a squared projection of the frequency vector $\bm{\omega}$ onto the eigenvector $\bm{v}^j$ that is scaled by inverse square of the associated eigenvalue $\lambda_j$. Thus, under the constraint of fixing the variance the frequency vector to $\sigma^2$, the collective behavior is strengthened by aligning the frequency vector $\bm{\omega}$ as close as possible with the most dominant eigenvectors (those associated with larger eigenvalues) and orthogonalizing $\bm{\omega}$ as best as possible to the least dominant eigenvectors (those associated with smaller eigenvalues). Thus, the optimal solution is obtained by setting $\bm{\omega}=\sigma\sqrt{N}\bm{v}^N$.

\begin{figure}[t]
\centering
\epsfig{file =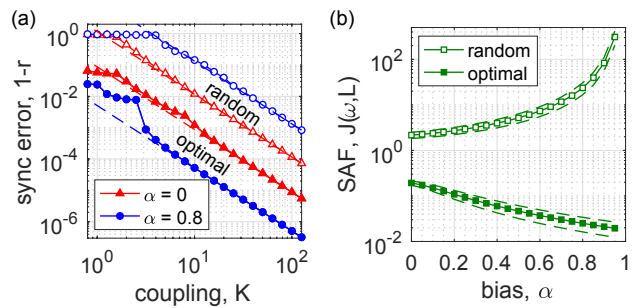, clip =,width=1.0\linewidth }
\caption{\emph{Optimal synchronization in networks with higher-order interactions.} (a) The synchronization error $1-r$ vs $K$ for random (open symbols) and optimal (closed symbols) frequencies for two choices of the bias parameter: $\alpha = 0$ (red triangles) and $0.8$ (blue circles), representing cases where interactions are exclusively defined by 1-simplexes and dominated by 2-simplexes, respectively, for a noisy geometric network (see text). (b) The Synchrony Alignment Function (SAF) $J(\bm{\omega},L)$ as a function of $\alpha$ for randomly-allocated (open squares) and optimal (closed squares) frequencies averages over $10^3$ networks.} \label{fig2}
\end{figure}

\section{Higher-Order Interactions Improve Collective Behavior for Optimized Systems}\label{sec:04}

In our first experiment, we highlight that random and optimized systems generally behave very differently, especially in the context of higher-order interactions. Specifically, we will show for optimized systems that collective behavior is improved by a stronger reliance on higher-order interactions, whereas it is diminished for random systems. Since simplicial complexes are geometrically embedded~\cite{Kaczynski2006} we consider a class of noisy geometric networks~\cite{Taylor2015} that contain both geometrically constrained and geometrically unconstrained edges between nodes uniformly placed on the unit disc in $\mathbb{R}^2$. With connected triangles, i.e., 2-simplexes, arising from geometrically constrained edges, we tune the prevalence of triadic interactions using a probability $p\in[0,1]$: (i) with probability $p$ each of the total $M=N\langle k^{(1)}\rangle/2$ edges is placed between the two closest nodes that are not yet connected and (ii) with probability $(1-p)$ each edge is placed randomly, where $\langle k^{(1)}\rangle$ is the target mean 1-simplex degree. Thus, $p$ tunes the prevalence of low-dimensional geometry in the network: in the limit $p\to1$ the network is purely geometric, while in the limit $p\to0$ the network is Erd\H{o}s-R\'{e}nyi~\cite{Erdos1960}. In Appendix~\ref{appB} we provide a more complete algorithm implementing the network model described above.

Taking one such network of size $N=500$ with mean degree $\langle k^{(1)}\rangle=10$ and $p=0.25$, we illustrate the effect of higher-order interactions on optimizing collective dynamics in Fig.~\ref{fig2}(a). We plot the synchronization error $1-r$ from direct simulations of Eq.~(\ref{eq:01}) as a function of $K$ for four cases, all under the constraint that the natural frequency vector has unit variance. First we consider the fully 1-simplex dominated case, i.e., $\alpha = 0$ so that coupling is purely dyadic, and plot the results for random and optimal choices of natural frequencies in open and closed red triangles, respectively. Note that the optimal choice of natural frequencies outperforms the random case, given by a set of natural frequencies drawn from the standard normal distribution, by about an order of magnitude. Next, we set $\alpha=0.8$, thereby strengthening higher-order interactions at the expense of pair-wise interactions, and plot the results for random and optimal choices of natural frequencies in open and closed blue circles, respectively. We note here that all simulations are done using Heun's method with a time step of $\Delta t =0.02$, intregrating over a transient of $5\times10^3$ timesteps and then averaged over a steady state of $2\times10^3$ time steps. We also plot the predicted synchronization error, given by $J(\bm{\omega},L)/2K^2$, for each case in dashed curves, which accurately capture the dynamics for sufficiently large coupling. 

This example highlights a critical feature of higher-order interactions in networks and their effect on collective dynamics. In particular, focusing on the optimal cases, the presence of higher-order interactions {\it improves} the optimal collective behavior supported by the system. Moreover, this phenomenon is generic: the more 2-simplex dominated a network is (i.e., the larger $\alpha$ is), the better the optimal states become. This is illustrated in Fig.~\ref{fig2}(b), where over an ensemble of $10^3$ networks built using the same parameters as the network using in Fig.~\ref{fig2}(a) we plot the value of the SAF as a function of the bias parameter $\alpha$ for randomly chosen frequencies and the optimal choice in open and closed squares, respectively. (The average over this ensemble is plotted with dashed curves indicating one standard deviation up and down.) Specifically, we see that as $\alpha$ increases, thus making the the network more 2-simplex dominated, the optimal state improves very smoothly and monotonically while the random states worsen. Thus, strengthening higher-order interactions in collective network dynamics not only improves the optimal states, but also widens the range of possible states that are supported.

\begin{figure}[t]
\centering
\epsfig{file =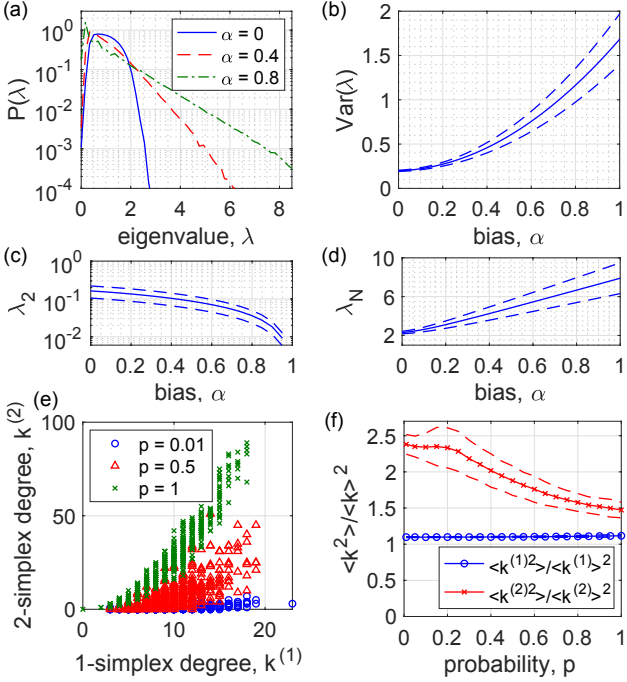, clip =,width=1.0\linewidth }
\caption{\emph{Spectral properties of a composite Laplacian.} (a) The eigenvalue spectrum $P(\lambda)$ of the composite Laplacian $L$ for $\alpha=0$ (solid blue), $0.4$ (dashed red), and $0.8$ (dot-dashed green) obtained from $10^3$ networks of size $N=500$ with mean degree $\langle k\rangle=10$ and $p=0.25$. (b) The variance of the eigenvalue spectrum along with the extremal eigenvalues (c) $\lambda_2$ and (d) $\lambda_N$ from the same ensemble. (e) 2-simplex degrees $k^{(2)}$ vs 1-simplex degrees $k^{(1)}$ for a single network realization and (f) the quantities $\langle k^{(1)2}\rangle/\langle k^{(1)}\rangle^2$ and $\langle k^{(2)2}\rangle/\langle k^{(2)}\rangle^2$ (blue circles and red crosses, respectively) obtained from an ensemble of $10^3$ networks as a function of the parameter $p$.} \label{fig3}
\end{figure}

\section{Broadening of Composite Laplacian Eigenspectrum Underlies Dichotomy For Optimized and Non-Optimized Systems}

To explain and further illustrate the improvement that occurs in collective network dynamics as a result of increased higher-order interactions, we investigate the spectral properties of the composite Laplacian $L=(1-\alpha)L^{(1)}+\alpha L^{(2)}$. Importantly, from Eq.~(\ref{eq:05}) we can see that while the structure of the eigenvectors of $L$ dictate the geometry of the optimal choice for the frequency vector $\bm{\omega}$, it is the eigenvalues that give insight into the quality of these optimal states. As an example, in Fig.~\ref{fig3}(a) we plot the eigenvalue spectrum of $L$ averaged across $10^3$ networks of size $N=500$ and built using the model described above with mean degree $\langle k^{(1)}\rangle=10$ and $p=0.25$ for $\alpha=0$ (solid blue), $0.4$ (dashed red), and $0.8$ (dot-dashed green). Note that as $\alpha$ increases and the higher-order interactions strengthen at the expense of pairwise interactions, the eigenvalue spectrum becomes broader. 

In fact, it is the broadening of the eigenvalue spectrum, and specifically the increase in the dominant eigenvalue $\lambda_N$, that corresponds to improving the optimal states, since, given the optimal choice $\bm{\omega}=\sigma\sqrt{N}\bm{v}^N$, we have $\|\bm{\theta}^*\|^2/N=J(\bm{\omega},L)/K^2=\sigma^2/(K\lambda_N)^2$. Here, we provide rigorous analytical insight on this mechanism by computing exactly the mean and variance of the eigenvalue spectrum in terms of moments of the various degrees using the trace of different powers of $L$. First, due to the conservation of the overall weighting of $L^{(1)}$ and $L^{(2)}$, the mean is always conserved to one: $\langle \lambda\rangle =N^{-1}\text{Tr}(L)=N^{-1}[(1-\alpha)\sum_{i}k_i^{(1)}/\langle k^{(1)}\rangle+\alpha\sum_{i}k_i^{(2)}/\langle k^{(2)}\rangle]=1$. Next, the variance $\text{Var}(\lambda)=\langle\lambda^2\rangle-\langle\lambda\rangle^2=N^{-1}\text{Tr}(L^2)-N^{-2}\text{Tr}^2(L)$ of the eigenvalue spectrum about this mean is given by
\begin{align}
\text{Var}(\lambda)&=(1-\alpha)^2\left(\frac{\langle k^{(1)2}\rangle}{\langle k^{(1)}\rangle^2}+\frac{1}{\langle k^{(1)}\rangle}\right)\nonumber\\&+2\alpha(1-\alpha)\left(\frac{\langle k^{(1)}k^{(2)}\rangle}{\langle k^{(1)}\rangle\langle k^{(2)}\rangle}+\frac{1}{2\langle k^{(1)}\rangle}\right)\nonumber\\&+\alpha^2\left(\frac{\langle k^{(2)2}\rangle}{\langle k^{(2)}\rangle^2}+\frac{\langle q\rangle}{4\langle k^{(2)}\rangle^2}\right)-1,\label{eq:06}
\end{align}
where $q_i=\sum_{j=1}^NA_{ij}^{(2)2}$. [See Appendix~\ref{appC} for the derivation of Eq.~(\ref{eq:06}).] In particular, varying $\alpha$ interpolates the variance between $\langle k^{(1)2}\rangle/\langle k^{(1)}\rangle^2+1/\langle k^{(1)}\rangle-1$ and $\langle k^{(2)2}\rangle/\langle k^{(2)}\rangle^2+\langle q\rangle/(4\langle k^{(2)}\rangle^2)-1$ in the extremes where connections are completely dominated by 1-simplex and 2-simplex coupling, respectively. Thus, when the latter form of the variance is larger, which we may expect when the 2-simplex degree distribution is more heterogeneous than the traditional 1-simplex degree distribution, strengthening higher-order interactions in turn broadens the eigenvalue spectrum of $L$. In general, as $\alpha$ is varied the eigenvalues interpolate between their respective values for $L^{(1)}$ and $L^{(2)}$, however their intermediate behavior is more complicated and left for future research.

In Fig.~\ref{fig3}(b), we plot the mean variance of the spectral density as a function of $\alpha$ which we calculated from the same ensemble as in panel (a) (indicating standard deviation with dashed curves). We observe a monotonic increase in the variance of the eigenvalue spectrum as higher-order interactions are strengthened, which is consistent with the broadening shown in panel (a). The extremal eigenvalues $\lambda_2$ and $\lambda_N$ follow this trend, decreasing and increasing, respectively, as illustrated in Figs.~\ref{fig3}(c) and (d). Moreover, we show over a full range of networks, from completely random to strongly geometric, that the 2-simplex degree distribution does in fact tend to be more heterogeneous than the 1-simplex degree distribution, thereby yielding improved collective dynamics as higher-order interactions are strengthened. In panel (e), we plot the 2- vs 1-simplex degrees for a single realization of the networks described above for parameters $p=0.01$, $0.5$, and $1$, representing random, partially geometric, and completely geometric cases. The concave-up trend for each case suggests that the 2-simplex degree distribution is in fact more heterogeneous than the 1-simplex degree distribution. For a more concrete picture, we plot in panel (f) the quantities $\langle k^{(1)2}\rangle/\langle k^{(1)}\rangle^2$ and $\langle k^{(2)2}\rangle/\langle k^{(2)}\rangle^2$ (in blue circles and red crosses, respectively) for the network model discussed above across a full range of the parameter $p$, representing completely random networks ($p\approx 0$) to completely geometric networks ($p\approx 1$). Each data point represent the mean over an ensemble of $10^3$ networks, with dashed curves representing one standard deviation. Here we see explicitly that over the full range we have that generically $\langle k^{(2)2}\rangle/\langle k^{(2)}\rangle^2>\langle k^{(1)2}\rangle/\langle k^{(1)}\rangle^2$, indicating that the phenomenon by which higher-order interactions improve optimal collective network dynamics in fact holds over a broad family of both random and geometric networks.

Furthermore, we may use this spectral analysis to shed light on more than just the optimal states, but also the worst possible state and random cases more broadly. First, analogous to the manner in which making the system more 2-simplex dominated increases $\lambda_N$, and in turn promotes optimal collective dynamics, the complementary decrease in $\lambda_2$ results in poorer worst-case collective dynamics, which would result in setting the frequency vector proportional to the first non-trivial eigenvector, $\bm{\omega}\propto\bm{v}^2$. Moreover, random frequency arrangements may be understood as follows. Constraining $\bm{\omega}$ to unit variance $\sigma^2$, it may be expanded using the eigenvector basis of $L$, $\bm{\omega}=\sum_{j=2}^Nc_j\bm{v}^j$, with $\sum_{j=2}^Nc_j^2=N\sigma^2$. Since heterogeneities are random and independent of network structure, the expected value of each coefficient is $E[c_j]=\pm\sqrt{N/(N-1)}$ thus and the expected value of $J(\bm{\omega},L)$ is given by
\begin{align}
E[J(\bm{\omega},L)]=\frac{1}{N}\sum_{j=2}^N\frac{N}{(N-1)\lambda_i^2}=\langle \lambda^{-2}\rangle,\label{eqn:07}
\end{align}
where the average is taken over all eigenvalues except for the trivial eigenvalue $\lambda_1=0$. When higher-order interactions are then strengthened, broadening the eigenvalue spectrum, the decrease of the smaller eigenvalues $\lambda_2,\lambda_3,\dots$, which tend towards zero [See Fig.~\ref{fig3}(a)] has a stronger effect on $\langle\lambda^{-2}\rangle$ than the increase of the larger eigenvalues $\dots,\lambda_{N-1},\lambda_N$. The overall effect of broadening the eigenvalue distribution by strengthening higher-order interactions is then increasing the expected value of the SAF and poorer expected collective behavior, even though, as we have seen above, the optimal states are improved.

\begin{figure}[t]
\centering
\epsfig{file =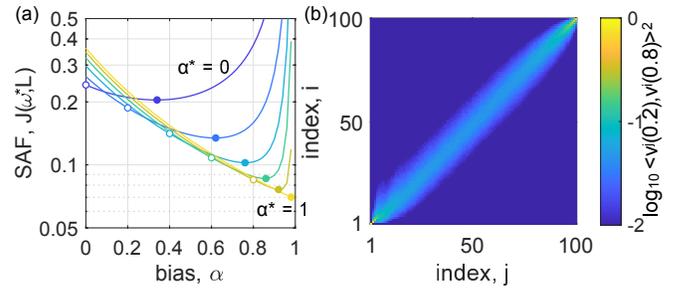, clip =,width=1.0\linewidth }
\caption{\emph{Geometric consistency in spectral properties of the composite Laplacian.} (a) For networks optimized at $\alpha^*=0,0.2,\dots,1$ (blue to yellow), the SAF $J(\bm{\omega}^*,L)$ of this solution as a function of $\alpha$ given the optimal frequency vector $\bm{\omega}^*$. Points along the curves at $\alpha=\alpha^*$ and the local minimum are denoted in open and closed circles, respectively. (b) The logarithm (base-10) of the squared projections $\langle\bm{v}^j(0.2),\bm{v}^i(0.8)\rangle^2$. Results are obtained from an ensemble of $10^3$ networks of size $N=100$ with mean degree $\langle k^{(1)}\rangle=10$ and $p=0.25$.} \label{fig4}
\end{figure}

\section{Geometric Consistency of Optimal Solutions}

Next we explore the robustness of optimal solutions as the bias parameter is varied. In particular, we find a geometric consistency in the optimal and near-optimal choices of $\bm{\omega}$ across a range of $\alpha$. To illustrate this phenomenon, we consider the optimal choices of $\bm{\omega}^*$ for a handful of given bias parameter values $\alpha^*=0,0.2,\dots,1$ and plot in Fig.~\ref{fig4}(a) these values of the SAF $J(\bm{\omega}^*,L)$ as a function of $\alpha$, with choices $\alpha^*=0$ and $1$ plotted in blue and yellow, respectively (and intermediate values interpolating these colors), for an ensemble of $10^3$ networks of size $N=100$ with mean degree $\langle k^{(1)}\rangle=10$ and $p=0.25$. 

First, we note that, although one may expect the minimum of these curves to occur at $\alpha=\alpha^*$ (denoted by the open circles), the improvement of optimal collective behavior as $\alpha$ increases causes this minimum to occur at another value $\alpha>\alpha^*$ (denoted by the closed circles). Thus, after optimizing a system at a particular bias $\alpha =\alpha^*$, increasing $\alpha$ results in improved collective behavior even without redesigning the frequency vector. Second, notice that for the case $\alpha^*=1$ (given by the yellow curve) the value of the SAF $J(\bm{\omega}^*,L)$ remains quite close to the true optimal $J(\bm{\omega},L)$ for all other $\alpha$. We see that this also generalizes: the SAF $J(\bm{\omega}^*,L)$ for a given value of $\alpha^*$ remains close to the optimal SAF $J(\bm{\omega},L)$ for all $\alpha<\alpha^*$. Thus, an optimal solution found for a $2$-simplex dominated networks remains consistently near-optimal as the network becomes more $1$-simplex dominated.

This phenomenon is due to a particular geometric property whereby the near-optimal subspaces of $\mathbb{R}^{N}$ for two different values of $\alpha$, namely the subspaces spanned by the eigenvectors $\bm{v}^j$ of $L$ with large eigenvalues $\lambda_j$ for the two values of $\alpha$ are largely overlapped. This is illustrated in Fig.~\ref{fig4}(b) for $\alpha = 0.2$ and $\alpha = 0.8$, where we plot the logarithm (base-10) of the squared projection $\langle\bm{v}^j(0.2),\bm{v}^i(0.8)\rangle^2$ for each $(i,j)$ pair, using the same ensemble of networks as above and denoting $\bm{v}^j(\alpha)$ as the $j^{\text{th}}$ eigenvector of $L$ for a bias parameter $\alpha$. Note the strong diagonal feature indicating a strong alignment of $\bm{v}^j(0.2)$ with $\bm{v}^i(0.8)$ for the corresponding eigenvector $i=j$ and other nearly adjacent eigenvectors $i\approx j$. This geometric consistency likely has to do with structure of clique complexes, namely that the 2-simplex structure is defined precisely by the 1-simplex structure and all non-zero entries of the matrix $A^{(2)}$ come from non-zero entries of $A^{(1)}$, and likely do not exist for more general cases of hypergraphs and simplicial complexes that are not clique complexes where 1- and 2-simplex structures may be uncorrelated.

\begin{figure}[t]
\centering
\epsfig{file =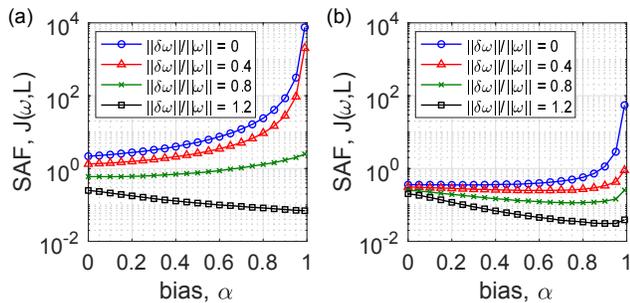, clip =,width=1.0\linewidth }
\caption{\emph{Constrained optimization.} The SAF $J(\bm{\omega},L)$ as a function of $\alpha$ obtained after optimal perturbations of sizes $\|\delta\bm{\omega}\|/\|\bm{\omega}\|=0$ (blue circles), $0.4$ (red triangles), $0.8$ (green crosses), and $1.2$ (black squares) are applied to a randomly drawn vector of frequencies (a) without and (b) with preprocessing the frequency vector using a (near) optimal permutation. Results are obtained from an ensemble of $10^2$ networks of size $N=100$ with mean degree $\langle k^{(1)}\rangle=10$ and $p=0.25$.} \label{fig5}
\end{figure}

\section{Constrained Optimization}

Lastly, we consider optimization of collective behavior in a more constrained scenario. Rather than just constraining the variance of a frequency vector $\bm{\omega}$ and allowing frequencies to be freely chosen otherwise (thereby allowing them to be aligned perfectly with a particular eigenvector), we assume a randomly chosen initial frequency vector is given and may only be modified by a perturbation of constrained size. Denoting this perturbation by $\delta\bm{\omega}=\bm{\omega}_{\text{new}}-\bm{\omega}$, we then constrain the relative size $\|\delta\bm{\omega}\|/\|\bm{\omega}\|$ while maintaining the variance of the frequency vector itself. This perturbation may be optimally designed in terms of the eigenvector expansion $\bm{\omega}=\sum_{j=2}^Nc_j\bm{v}^j$ by orthogonalizing away from the eigenvectors with smallest associated eigenvalues in order to eliminate the largest contributions to the SAF. To do this, we let $\delta\bm{\omega}=\sum_{j=2}^N\beta_j\bm{v}^j$ and, for as large $k$ as possible, let $\beta_j=-c_j$ for $j=2,\dots,k$, $\beta_{j}=c_j$ for $j=k+1,\dots,N-1$, and $\beta_N=c_N(\sqrt{1+\sum_{j=2}^kc_j^2/c_N^2}-1)$, resulting in $\bm{\omega}_{\text{new}}=\sum_{j=2}^N\gamma_j\bm{v}^j$ with $\gamma_j=0$ for $j=2,\dots,k$, $\gamma_j=c_j$ for $j=k+1,\dots,N-2$, and $\gamma_N=c_N\sqrt{1+\sum_{j=2}^kc_j^2/c_N^2}$. Note that this both orthogonalizes $\bm{\omega}_{\text{new}}$ against the eigenvectors with smallest eigenvalues while increasing the alignment with $\bm{v}_N$ in order to conserve the variance of $\bm{\omega}$. In Fig.~\ref{fig5}(a), we plot the resulting SAF $J(\bm{\omega},L)$ averaged over $10^2$ networks from an ensemble using the same parameters as in Figs.~\ref{fig2} and \ref{fig3} after imposing such a perturbation of sizes $\|\delta\bm{\omega}\|/\|\bm{\omega}\|=0$, $0.4$, $0.8$, and $1.2$ (plotted in blue circles, red triangles, green crosses, and black squares) to a random frequency vector with normally distributed entries. Note that the maximum possible perturbation that conserves the standard deviation of the frequencies is $\|\delta\bm{\omega}\|/\|\bm{\omega}\|=2$, obtained by $\bm{\omega}_{\text{new}}=-\bm{\omega}$. 

Another realistic possibility  is that, before a perturbation is applied, the frequencies are (nearly) optimally rearranged to obtain a permutation of the initially given frequency vector. Here we obtain such a permutation using a simple accept-reject algorithm that interchanges randomly chosen pairs of frequencies if the exchange decreases the SAF. This preprocessing technique allows for more efficient perturbations, as we see in Fig.~\ref{fig5}(b), which improve upon the results in Fig.~\ref{fig5}(a). In particular, in such a constrained optimization scenario, we observe that there is often an ideal balance of dyadic to triadic interactions, i.e., a critical value of $\alpha$ that lies between zero and one, for a given perturbation size. 

This phenomenon can be viewed as a combination between the cases of random frequencies, where higher-order interactions impede collective, and the case of optimal (freely-tunable) frequencies, where higher-order interactions improve collective behavior. Specifically, the presence of a constraint allows higher-order interactions to improve the constrained optimal states, but only to a certain point since the purely optimal choice of frequencies is unattainable, as the frequency vector cannot be precisely aligned with the eigenvector vector $\bm{v}^N$.

\section{Discussion}

Given the role of collective behavior in the function of physical, biological, and neurological systems where higher-order interactions may play a critical role in shaping system dynamics, understanding how higher-order interactions balance with dyadic interactions and affect collective behavior is an important question for a wide range of disciplines and applications. In this paper, we have addressed the topic of optimization for collective behavior in networks with higher-order interactions, focusing on clique complexes, and found that as higher-order interactions are equitably strengthened relative to dyadic interactions, optimal collective behavior improves. This phenomenon stems from the broadening of the eigenvalue spectrum of a composite Laplacian matrix that encodes the collective dynamics and network structure at multiple orders and generalizes the Synchrony Alignment Function framework to this important case. In particular, as the spectrum broadens the dominant eigenvalue(s) increase, which leads to this improvement. Moreover, we find that optimal solutions are robust over different balances between the relative strengths of  dyadic and triadic interactions and that the broadening of the eigenvalue spectrum also widens the range of possible collective states supported by the network. We also find in more tightly constrained optimization scenarios that an ideal balanced between dyadic and triadic interactions occurs at a nontrivial, critical value of the bias parameter for networks to support the strongest possible collective behavior. Interestingly, the improvement of optimal collective behavior stemming from the broadening of the eigenvalue spectrum for the case of heterogeneous dynamical units lies in contrast to the case of identical units, where optimal networks stem from the concentration of the non-trivial eigenvalue spectrum~\cite{Barahona2002PRL}.

These results shed light on the question of why higher-order interactions may be important in various applications that exhibit collective behavior. In particular, by modifying one's balance between dyadic and higher-order interactions, a system may self-regulate not only to modify (improve or worsen) its current collective state, but also broaden or contract the set of possible collective states that the system may support under further modification of the individual unit's local dynamics (here given by the oscillators' natural frequencies). This observation may be particularly useful for generating hypotheses for systems that exhibit a strong tendency for re-organization and self-regulation such as the brain where empirical evidence suggests that higher-order interactions play a role in collective behavior~\cite{Yu2011Neuro,Petri2014Interface,Giusti2016JCN,Reimann2017,Sizemore2018JCN} and an optimal range of dynamic behavior is crucial for function~\cite{Kinouchi2006NatPhys,Larremore2011PRL}.

This work introduced a composite Laplacian matrix that encodes network structure at multiple orders to generalize the SAF framework for optimizing collective behavior~\cite{Skardal2014PRL}. The SAF framework has already been generalized for a number of scenarios involving optimization, including optimization with directed interactions~\cite{Skardal2016Chaos}, finding optimal network perturbations~\cite{Taylor2016SIAM}, optimizing networks with chaotic oscillators~\cite{Skardal2017Chaos}, addressing uncertainty in local dynamics~\cite{Skardal2019SIAM}, and uncovering geometric unfolding of networks~\cite{Arola2021Chaos}. Future work may address many of these generalizations in the context of networks with higher-order interactions. Moreover, since we have restricted our attention to the case of clique complexes, it remains an open question how our findings would change if one were to consider other network structures such as more general simplicial complexes, hypergraphs, and non-geometric networks.

\acknowledgements

PSS acknowledges support from NSF grant MCB-2126177. DT acknowledges support from NSF grant DMS-2052720 and Simons Foundation award \#578333. L.A.-F. and A.A. acknowledge support from the Spanish MINECO (Grant No. PGC2018-094754-B-C2). A.A. also acknowledges support from the Generalitat de Catalunya (Grant No. 2017SGR-896), the Universitat Rovira i Virgili (Grant No. 2017PFR-URV-B2-41), and the ICREA Academia and the James S. McDonnell Foundation (Grant No. 220020325). The authors thank Giulio Burgio for  fruitful discussions.

\appendix

\section{Additional Higher-Order Coupling Terms}\label{appA}

In addition to the higher-order coupling term in Eq.~(\ref{eq:01}), phase-reduction analyses of limit-cycle oscillators~\cite{Ashwin2016PhysD,Leon2019PRE}, both with linear and nonlinear coupling, point to the additional possible coupling term, given by
\begin{align}
\frac{1}{4\langle k^{(2)}\rangle}\sum_{j=1}^N\sum_{l=1}^NB_{ijl}\sin(\theta_j+\theta_l-2\theta_i).\label{eq:mat:a}
\end{align}
Linearizing Eq.~(\ref{eq:mat:a}) yields
\begin{align}
&-\left(k_i^{(2)}\theta_i-\frac{1}{4}\sum_{j=1}^NA_{ij}\left(\sum_{l=1}^NA_{jl}A_{li}\right)\theta_j\right.\nonumber\\
&~~~~~~~~~~~~~~~~~~~~~\left.-\frac{1}{4}\sum_{j=1}^NA_{ji}\left(\sum_{l=1}^NA_{il}A_{lj}\right)\theta_j\right)/\langle k^{(2)}\rangle,\label{eq:mat:b}
\end{align}
which can be expressed in vector form as $L^{(2)'}=(D^{(2)}-\left(A^{(2)}/4+A^{(2)T}/4\right))/\langle k^{(2)}\rangle$. Note, however, that in the case of undirected, unweighted networks where $A^{(2)}=A^{(2)T}$ this reduces to $L^{(2)'}=(D^{(2)}-A^{(2)}/2)/\langle k^{(2)}\rangle=L^{(2)}$, which was defined just below Eq.~(\ref{eq:02}). Thus, after linearization the higher-order interactions given in Eq.~(\ref{eq:mat:a}) are interchangeable with those studied in the main text and given in Eq.~(\ref{eq:01}).

\section{Algorithm for the Clustered Network Model}\label{appB}

Here we present a detailed algorithm for generating networks from the model described in the main text in Sec.~\ref{sec:04}. We begin with three main input parameters: (i) the networks size, i.e., number of nodes, $N$, (ii) the mean 1-simplex degree $\langle k^{(1)}\rangle$, and (iii) the clustering, i.e., triadic interaction, parameter $p$. Given $N$ nodes, we seek to place a total of $M=\langle k\rangle N/2$ total links between nodes to ensure the mean 1-simplex degree is attained.

As a first step, we place $N$ nodes uniformly within the unit disc. We then proceed to place links of the first kind, specifically $pM$ links that connect nearby nodes. We evaluate the distance between each pair nodes that are not connected and place a link between the pair with the shortest distance. We then iteratively repeat this process until each of the $pM$ such links are placed. We then move on to links of second kind, specifically $(1-p)M$ random links. We then consider all pairs of unconnected nodes and connect one such pair at random. Again, we iteratively repeat this process until each of the $(1-p)M$ such links are placed.

Having the 1-simplex structure. i.e., the adjacency matrix $A$ completed, we then move on to the 2-simplex structure. Specifically, this depends on thje 1-simplex structure, so to populate the adjacency tensor $B$ we simply do an exhaustive scan through the entire network, identifying each distinct triplet $(i,j,k)$ that form a triangle, i.e., have $A_{ij},A_{jl},A_{li}=1$, and fill in the appropriate entries $B_{ijl},B_{jil},B_{ilj},B_{jli},B_{lij},B_{lji}=1$, completing the 2-simplex structure of the network.

\section{Derivation of Eq.~(\ref{eq:06})}\label{appC}

We begin by writing the variance simply as
\begin{align}
\text{Var}(\lambda)=\langle \lambda^2\rangle - \langle\lambda\rangle^2=N^{-1}\text{Tr}(L^2) - N^{-2}\text{Tr}(L)^2.\label{eq:mat:00}
\end{align}
Note that since $\text{Tr}(L)=N$, we have that the last term reduces to one. Thus, our focus turns to the quantity $\text{Tr}(L^2)$. First, using the fact that the network is undirected, and therefore $A^{(2)}=A^{(2)T}$ we write
\begin{align}
L^2&=\left[(1-\alpha)\frac{D^{(1)}-A^{(1)}}{\langle k^{(1)}\rangle}+\alpha\frac{D^{(2)}-A^{(2)}/2}{\langle k^{(2)}\rangle}\right]^2\nonumber\\
&= (1-\alpha)^2\frac{D^{(1)2}-D^{(1)}A^{(1)}-A^{(1)}D^{(1)}+A^{(1)2}}{\langle k^{(1)}\rangle^2}\nonumber\\
&+ \alpha(1-\alpha)\frac{D^{(1)}D^{(2)}-D^{(1)}A^{(2)}/2-A^{(1)}D^{(2)}+A^{(1)}A^{(2)}/2}{\langle k^{(1)}\rangle\langle k^{(2)}\rangle}\nonumber\\
&+ \alpha(1-\alpha)\frac{D^{(2)}D^{(1)}-D^{(2)}A^{(1)}-A^{(2)}D^{(1)}/2+A^{(2)}A^{(1)}/2}{\langle k^{(1)}\rangle\langle k^{(2)}\rangle}\nonumber\\
&+\alpha^2 \frac{D^{(2)2}-D^{(2)}A^{(2)}/2-A^{(2)}D^{(2)}/2+A^{(2)2}/4}{\langle k^{(2)}\rangle^2}.\label{eq:mat:01}
\end{align}
Next, applying the trace to Eq.~(\ref{eq:mat:01}) and rearranging yields
\begin{align}
\text{Tr}(L^2)&=\frac{(1-\alpha)^2}{\langle k^{(1)}\rangle^2}\left[\text{Tr}(D^{(1)2})-2\text{Tr}(D^{(1)}A^{(1)})+\text{Tr}(A^{(1)2})\right]\nonumber\\
&+ \frac{2\alpha(1-\alpha)}{\langle k^{(1)}\rangle\langle k^{(2)}\rangle}\left[\text{Tr}(D^{(1)}D^{(2)})-\text{Tr}(D^{(1)}A^{(2)})/2\right.\nonumber\\&~~~~~~~~~~~~~~~~~~~~~~~~~~\left.-\text{Tr}(A^{(1)}D^{(2)})+\text{Tr}(A^{(1)}A^{(2)})/2\right]\nonumber\\
&+\frac{\alpha^2}{\langle k^{(2)}\rangle^2} \left[\text{Tr}(D^{(2)2})-2\text{Tr}(D^{(2)}A^{(2)})/2+\text{Tr}(A^{(2)2})/4\right].\label{eq:mat:02}
\end{align}
Since no self-links exist and triangles only exist between three distinct nodes, we have that $A_{ii}^{(1)}=A_{ii}^{(2)}=0$ for $i=1,\dots,N$, so that $\text{Tr}(D^{(1,2)}A^{(1,2)})=\sum_{i=1}^ND_{i}^{(1,2)}A_{ii}^{(1,2)}=0$ implying that each mixed term in Eq.~(\ref{eq:mat:02}) vanishes, simplifying to
\begin{align}
\text{Tr}(L^2)&=\frac{(1-\alpha)^2}{\langle k^{(1)}\rangle^2}\left[\text{Tr}(D^{(1)2})+\text{Tr}(A^{(1)2})\right]\nonumber\\
&+ \frac{2\alpha(1-\alpha)}{\langle k^{(1)}\rangle\langle k^{(2)}\rangle}\left[\text{Tr}(D^{(1)}D^{(2)})+\text{Tr}(A^{(1)}A^{(2)})/2\right]\nonumber\\
&+\frac{\alpha^2}{\langle k^{(2)}\rangle^2} \left[\text{Tr}(D^{(2)2})+\text{Tr}(A^{(2)2})/4\right].\label{eq:mat:03}
\end{align}
The traces of each of the matrices $D^{(1)2}$, $D^{(1)}D^{(2)}$, and $D^{(2)2}$ are given simply by
\begin{align}
\text{Tr}(D^{(1)2})&=\sum_{i=1}^Nk_i^{(1)2},\label{eq:mat:04}\\
\text{Tr}(D^{(1)}D^{(2)})&=\sum_{i=1}^Nk_i^{(1)}k_i^{(2)},\label{eq:mat:05}\\
\text{Tr}(D^{(2)2})&=\sum_{i=1}^Nk_i^{(2)2},\label{eq:mat:06}
\end{align}
while the traces of each of the matrices $A^{(1)2}$, $A^{(1)}A^{(2)}$, and $A^{(2)2}$ are given by
\begin{align}
\text{Tr}(A^{(1)2})&=\sum_{i=1}^N\left(\sum_{j=1}^NA^{(1)}_{ij}A^{(1)}_{ji}\right)=\sum_{i=1}^Nk_i^{(1)}\label{eq:mat:07}\\
\text{Tr}(A^{(1)}A^{(2)})&=\sum_{i=1}^N\left(\sum_{j=1}^NA^{(1)}_{ij}A^{(2)}_{ji}\right)=\sum_{i=1}^Nk_i^{(2)}\label{eqmat:08}\\
\text{Tr}(A^{(2)2})&=\sum_{i=1}^N\left(\sum_{j=1}^NA^{(2)}_{ij}A^{(2)}_{ji}\right)=\sum_{i=1}^Nq_i\label{eq:mat:09}
\end{align}
where we have used that $A^{(1)}$ and $A^{(2)}$ are undirected, $A^{(1)}$ is unweighted, and $q_i=\sum_{j=1}^NA^{(2)2}_{ij}$. Finally, inserting Eqs.~(\ref{eq:mat:04}) and (\ref{eq:mat:09}) into Eq.~(\ref{eq:mat:03}) and dividing by $N$ yields
\begin{align}
N^{-1}\text{Tr}(L^2)&=(1-\alpha)^2\left(\frac{\langle k^{(1)2}\rangle}{\langle k^{(1)}\rangle^2}+\frac{1}{\langle k^{(1)}\rangle}\right)\nonumber\\&+2\alpha(1-\alpha)\left(\frac{\langle k^{(1)}k^{(2)}\rangle}{\langle k^{(1)}\rangle\langle k^{(2)}\rangle}+\frac{1}{2\langle k^{(1)}\rangle}\right)\nonumber\\&+\alpha^2\left(\frac{\langle k^{(2)2}\rangle}{\langle k^{(2)}\rangle^2}+\frac{\langle q\rangle}{4\langle k^{(2)}\rangle^2}\right).\label{eq:mat:10}
\end{align}
Inserting Eq.~(\ref{eq:mat:10}) into Eq.~(\ref{eq:mat:00}) recovers Eq.~(\ref{eq:06}), as desired.

\bibliographystyle{plain}

\end{document}